\documentclass[sigconf]{acmart}
\usepackage{algorithm}
\usepackage{algorithmicx}
\usepackage{algpseudocodex}
\usepackage{multirow}
\usepackage{subcaption}
\usepackage{xspace}

\usepackage{array,tabularx,booktabs}
\newcolumntype{L}[1]{>{\raggedright\arraybackslash}p{#1}}
\newcolumntype{Y}{>{\raggedright\arraybackslash}X}

\AtBeginDocument{%
  }

\setcopyright{cc}
\copyrightyear{2026}
\acmYear{2026}
\setcctype{by}
\acmConference[SeQureDB '26]{Workshop on Secure and Private Data Management}{May 31-June 05, 2026}{Bengaluru, India}
\acmBooktitle{Workshop on Secure and Private Data Management (SeQureDB '26), May 31-June 05, 2026, Bengaluru, India}
\acmDOI{10.1145/3807894.3810270}
\acmISBN{979-8-4007-2219-6/2026/05}

\begin{document}

\title{Query Cost Model Calibration in Confidential Virtual Machines}

\author{Qihan Zhang}
\orcid{0009-0005-5785-8766}
\affiliation{%
  \institution{University of Southern California}
  \city{Los Angeles}
  \country{USA}
}
\email{qihanzha@usc.edu}

\author{Mengyuan Li}
\orcid{0009-0008-2721-4021}
\affiliation{%
  \institution{University of Southern California}
  \city{Los Angeles}
  \country{USA}
}
\email{mli49061@usc.edu}

\author{Ibrahim Sabek}
\orcid{0009-0006-2102-5241}
\affiliation{%
  \institution{University of Southern California}
  \city{Los Angeles}
  \country{USA}
}
\email{sabek@usc.edu}

\renewcommand{\shortauthors}{Zhang et al.}
\newcommand{\oursystem}{\textit{CM Calibration}}
\begin{abstract}

With the growing adoption of Confidential Computing, running databases in confidential virtual machines (CVMs) such as AMD SEV-SNP has become an attractive way to protect sensitive cloud data with minimal changes to legacy DBMSs. However, analytical queries in such CVMs often suffer substantial overhead, and prior database work has largely stopped at benchmarking these slowdowns rather than optimizing them. We show that this problem stems from a hardware--software mismatch: query optimizers still rely on KVM-oriented (non-encrypted VM) cost assumptions that no longer hold in CVMs. To address this, we propose a lightweight CVM-aware cost calibration. It models two dominant sources of optimizer-facing overhead: data-movement and RMP-related translation using simple physical proxies already available to the optimizer. Experiments show that the calibration significantly narrows the KVM/CVM performance gap, recovering up to 48\% performance and even outperforming the KVM baseline on some workloads.
\end{abstract}

\maketitle
\section{Introduction}
\label{section:introduction}

As cloud infrastructure increasingly hosts critical database workloads on third-party platforms~\cite{armbrust2010view,dageville2016snowflake,verbitski2017amazon}, protecting sensitive data without sacrificing performance has become a central challenge~\cite{chow2009controlling,fuller2017sok,naehrig2011can}. Confidential computing addresses this by protecting data in use within Trusted Execution Environments (TEEs). TEEs broadly fall into two classes: process-based designs such as Intel SGX~\cite{costan2016intel}, which offer strong isolation but often require intrusive application changes, and VM-based designs such as AMD SEV-SNP~\cite{sev2020strengthening}, which protect entire virtual machines with better compatibility but a larger trusted software stack. We focus on the latter, namely Confidential Virtual Machines (CVMs), such as AMD SEV-SNP, because they let legacy DBMSs like PostgreSQL~\cite{postgres} run with no source-code change, making them attractive for practical deployment~\cite{mulligan2021confidential,misono2024confidential,atiiq2024demystifying,amdsevnocodechangeslegacydb,intel_tdx_no_modifications}. Throughout the paper, we use \emph{CVM} to denote a hardware-protected confidential VM, and \emph{KVM} to denote the conventional unencrypted VM baseline.

However, this convenience comes with non-trivial overheads in memory protection, address translation, and I/O~\cite{misono2024confidential,akram2021performance}. Prior studies on DBMSs in CVMs mainly focus on benchmarking these penalties~\cite{maliszewski2021price,lutsch2024benchmarking,lutsch2025analysis,battiston2024duckdb,qiu2024price}. While they show that CVMs can substantially slow down database workloads, they do not examine how such overhead distorts core DBMS components, especially the query optimizer. Since optimizers choose among scan paths, join methods, and execution structures by comparing estimated costs~\cite{selinger1979access,graefe1993volcano,graefe1995cascades}, a CVM-unaware cost model can misrank operators and produce poor plans. This creates a hardware--software mismatch: cost estimations for KVMs no longer hold in CVMs.

Focusing on AMD SEV-SNP, we attribute this mismatch to two architectural properties of CVM (empirically evaluated in Section~\ref{section:motivation}). First, data movement in a CVM more often involves encryption-related processing and extra copying across the storage--memory hierarchy. This gives rise to \textit{data movement overhead}, including bounce-buffer copying on I/O paths and the higher cost of transferring large volumes of encrypted data under memory- or bandwidth-intensive execution~\cite{misono2024cvmexplained,mofrad2018comparison}. Second, SEV-SNP introduces the Reverse Map Table (RMP), a hardware metadata structure that validates page ownership and access permissions during address translation (i.e., mapping virtual addresses to physical pages). This gives rise to \textit{RMP overhead}, which raises the cost of memory-access paths triggering page-table walks and metadata checks, making random-access operators such as index scans and probe-heavy nested-loops more expensive~\cite{sev2020strengthening,amd56860snpabi,wilke2020sevurity,wang2025revisiting}. These penalties are highly non-uniform across operators, so plans that are reasonable in KVM can become poor choices in SEV-SNP. To address this problem, we propose a lightweight CVM-aware calibration centered on the query optimizer’s \emph{cost model}. Rather than redesigning the execution engine, it adjusts operator costing so that the DBMS better reflects the true relative prices of scan paths and join strategies in SEV-SNP. The key idea is to use simple physical proxies already exposed to the optimizer, such as pages touched, tuple volume, and memory usage, to inject targeted penalties for RMP-sensitive random access and data movement. This preserves the optimizer’s architecture while making plan decisions more faithful to CVMs. This calibration introduces essentially negligible overhead, as it only lightly modifies the optimizer’s cost estimation logic without adding new execution-time mechanisms.

We implement the calibration in PostgreSQL and evaluate it on standard OLAP benchmarks (JOB~\cite{leis2015good}, CEB~\cite{flowloss}, Stack~\cite{marcus2021bao}, and TPC-DS~\cite{tpcds}), saving at most 48\% execution time in CVMs. In summary, our contributions are:

\par\noindent\hangindent=1.6em\hangafter=1(1)\hspace{0.15em}
We show that RMP and data movement overheads explain the main optimizer-facing performance distortions relevant to query planning. Considering them in the cost model can improve DBMS performance in CVMs.

\par\noindent\hangindent=1.6em\hangafter=1(2)\hspace{0.15em}
We propose a CVM-calibrated cost model, which is a practical CVM-aware cost-model calibration extension that significantly narrows the KVM/CVM performance gap and can even outperform the KVM baseline.

\par\noindent\hangindent=1.6em\hangafter=1(3)\hspace{0.15em}
We implement the CVM-aware calibration inside PostgreSQL for CVMs, recovering up to 48\% performance and even outperforming the non-encrypted baseline on some workloads.

\section{Motivation}
\label{section:motivation}

SEV-SNP introduces at least two optimizer-relevant penalties, both rooted in its security mechanisms. The first is \emph{data movement cost}. In a CVM, transfers along the storage--memory path are no longer plain copies: they typically involve encrypted guest memory, additional I/O copying, and synchronization through intermediate software or hardware layers. A key mechanism is the \emph{bounce buffer}, a shared staging area used when data cannot be moved directly between devices and protected guest memory~\cite{misono2024cvmexplained}. Since different query plans stress this path differently, the resulting overhead is strongly plan-dependent. The second is \emph{RMP-check cost}. SEV-SNP protects guest memory with additional metadata and validation, commonly tied to the Reverse Map Table (RMP)~\cite{sev2020strengthening,wilke2020sevurity}. On translation-related events such as TLB or cache misses, the processor may perform page-table walks and protected-memory checks before completing the access. As a result, SEV-SNP can raise address-translation and memory-access latency even when the query plan is identical between KVM and CVM.

\begin{figure}[h]
  \centering
  \includegraphics[width=0.3\textwidth]{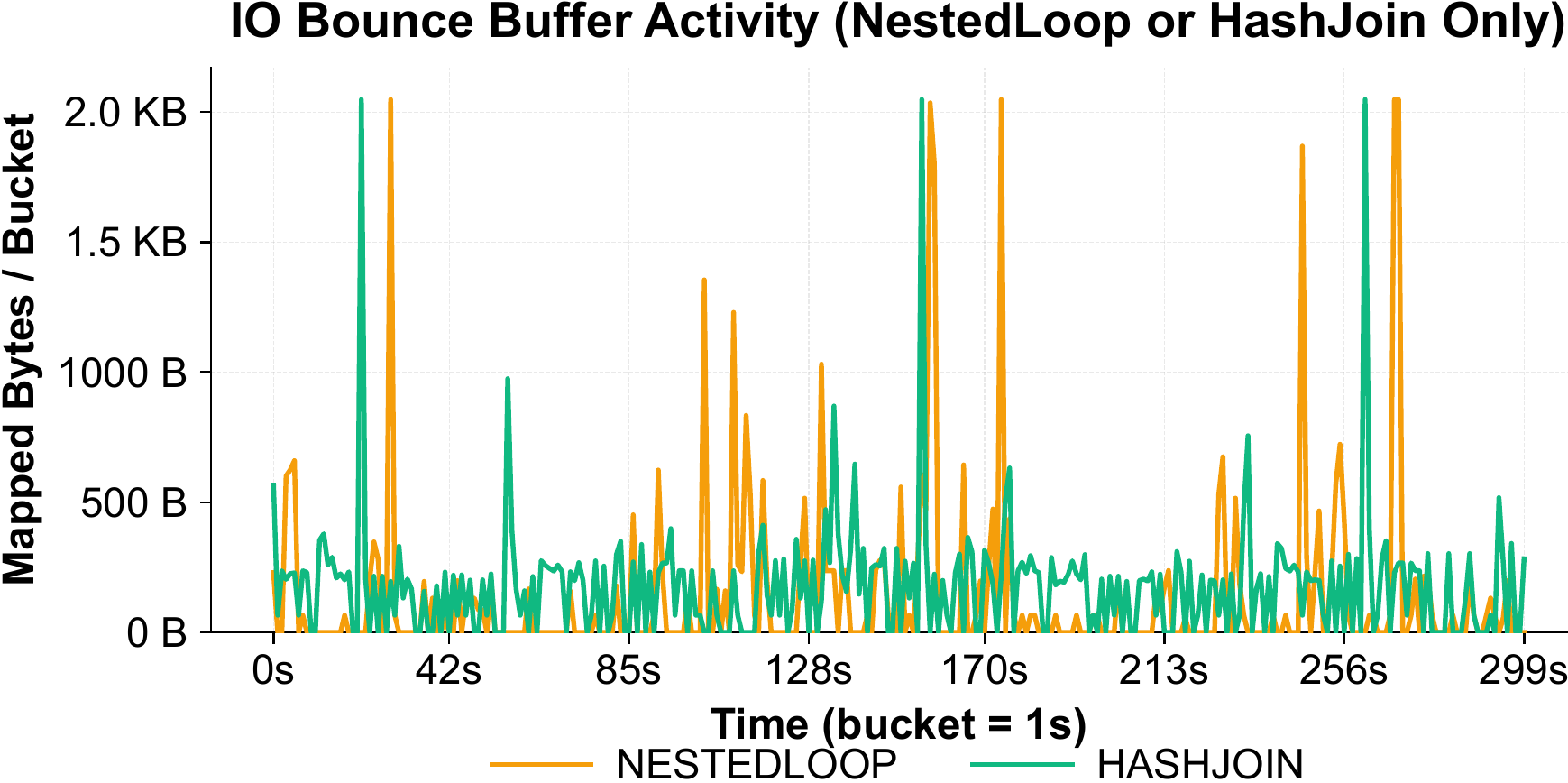}
  \vspace{-5pt}
 \caption{\small Evidence of \emph{data movement cost} in SEV-SNP. Hash join induces denser bounce-buffer activity, while nested-loop join induces less frequent but burstier activity.}
  \label{fig:bouncebuffer_activities}
\end{figure}

\noindent\textbf{Experimental Evidence.}
Figure~\ref{fig:bouncebuffer_activities} and Table~\ref{tab:rmp_evidence_17a} show that both penalties are observable. To expose the \emph{data movement cost}, we repeatedly run JOB ~\cite{leis2015good} in SEV-SNP under two settings, \emph{hash-join-only} and \emph{nested-loop-only} (disable other join strategies), with all other settings unchanged. For each setting, we monitor bounce-buffer activity for 300 seconds, partition the trace into 1-second buckets, and record transferred bytes (\emph{mapped bytes}) per bucket. The figure reveals clear operator-dependent movement patterns: hash join yields denser, more continuous activity, consistent with sustained movement during scans and hash build/probe phases, whereas nested loop is less frequent but more bursty, matching its operator behavior. This shows that SEV-SNP overhead is not a fixed slowdown but depends on how a plan moves data, motivating a \emph{data movement cost} in the operator cost model. To expose the \emph{RMP-check cost}, we run the \emph{same SQL (JOB 17a) under the same query plan} in KVM and SEV-SNP and collect low-level counters. The key signal is CPI (Cycles Per Instruction) inflation in SEV-SNP: instructions grow by only $1.23\times$, while cycles grow by $2.25\times$, raising CPI from $1.28$ to $2.33$. This indicates that SEV-SNP mainly increases the \emph{cost per instruction} rather than the amount of work performed. The pattern is consistent with translation-related slowdown: TLB misses trigger page-table walks, and protected-page accesses may incur extra RMP validation. Miss metrics do not indicate a miss-rate explosion. dTLB misses per cycle rise only modestly, as expected, since the same plan should preserve broadly similar data-access locality across KVM and CVM. iTLB misses per cycle increase more noticeably, suggesting higher instruction-side translation cost under RMP checks. Cache misses per cycle slightly decrease, which is also compatible with SEV-SNP-induced stall amplification: when total cycles grow substantially, longer stall time per miss can dilute miss rates normalized by cycle. Overall, the evidence is consistent with \emph{latency amplification on existing translation and memory-access paths}, confirming the existence of RMP-check cost.

\begin{table}[t]
\centering
\small
\setlength{\tabcolsep}{4pt}
\caption{\small Evidence of \emph{RMP-check cost} in SEV-SNP. For the same SQL and plan, SEV-SNP shows clear CPI inflation relative to KVM.}
\label{tab:rmp_evidence_17a}
\begin{tabular}{@{}lrrr@{}}
\toprule
\textbf{Metric} & \textbf{KVM} & \textbf{SEV-SNP} & \textbf{$\times$ (SNP/KVM)} \\
\midrule
Exec time (s)                         & 14.85 & 18.26 & 1.23 \\
Cycles ($\times 10^{11}$)             & 1.22  & 2.75   & 2.25 \\
Instructions ($\times 10^{11}$)       & 0.96  & 1.18   & 1.23 \\
CPI (cycles/inst)                     & 1.28  & 2.33   & 1.82 \\
dTLB misses/cycle ($\times 10^{-4}$)  & 1.51  & 1.93   & 1.28 \\
iTLB misses/cycle ($\times 10^{-3}$)  & 1.11  & 3.30   & 2.98 \\
Cache misses/cycle ($\times 10^{-3}$) & 2.65  & 2.09   & 0.79 \\
\bottomrule
\end{tabular}
\end{table}

\section{Overview of CVM-calibrated Cost Model}
\label{section:cost_model}

\begin{table*}[t]
\centering
\caption{a light-weight CVM-aware calibration for main operators.}
\vspace{-5pt}
\label{tab:operator_cost_model}
\small
\setlength{\tabcolsep}{6pt}
\renewcommand{\arraystretch}{1.18}
\begin{tabularx}{\textwidth}{@{}L{0.12\textwidth}Y L{0.12\textwidth}Y@{}}
\toprule
\textbf{Operator} & \textbf{Formula (Base cost with CVM overhead)} & \textbf{Operator} & \textbf{Formula (Base cost with CVM overhead)} \\
\midrule

Sequential scan (SS) &
$C_{\textsc{SS}}(R)=C^{\textsc{base}}_{\textsc{Seq}}(R)\allowbreak+\textsc{datamovecost}\!\big(ws^{\textsc{SS}}(R),x(R)\big)$ &
Bitmap scan (BS) &
$C_{\textsc{BS}}(R)=C^{\textsc{base}}_{\textsc{BS}}(R)\allowbreak
+\textsc{datamovecost}\!\big(ws^{\textsc{BS}}(R),x(R)\big)$ \\
\addlinespace[2pt]

Index (only) scan [I(O)S] &
$C_{\textsc{I(O)S}}(R,I)=C^{\textsc{base}}_{\textsc{I(O)S}}(R,I)\allowbreak
+\textsc{rmptcost}\!\big(rows_{\textsc{out}}(R)\!\cdot\!H(I),x(R,I)\big)
+\textsc{datamovecost}\!\big(ws^{\textsc{I(O)S}}(R,I),x(R,I)\big)$ &
Nested-loop join (NL) &
$C_{\textsc{NL}}(R,S)=C^{\textsc{base}}_{\textsc{NL}}(R,S)\allowbreak
+\textsc{datamovecost}\!\big(ws^{\textsc{NL}}(R,S),x(R)\big)
+\mathbb{I}_{\textsc{idx}}(R,S)\cdot \textsc{rmptcost}\!\big(rows(R)\!\cdot\!H(I_S),x(R)\big)$ \\
\addlinespace[2pt]

Hash join (HJ) &
$C_{\textsc{HJ}}(R,S)=C^{\textsc{base}}_{\textsc{HJ}}(R,S)\allowbreak
+\textsc{datamovecost}\!\big(ws^{\textsc{HJ}}(S),x_{\textsc{HJ}}(S)\big)
+\textsc{rmptcost}\!\big(a_{\textsc{HJ}}(R,S),x_{\textsc{HJ}}(S)\big)$ &
Merge join (MJ) &
$C_{\textsc{MJ}}(R,S)=C^{\textsc{base}}_{\textsc{MJ}}(R,S)\allowbreak
+\textsc{datamovecost}\!\big(ws^{\textsc{MJ}}(R,S),x(R,S)\big)
+\mathbb{I}_{\textsc{idx}}(R,S)\cdot \textsc{rmptcost}\!\big(a_{\textsc{MJ}}(R,S),x(R,S)\big)$ \\
\addlinespace[2pt]

Materialize (MAT) &
$C_{\textsc{Mat}}(R)=C^{\textsc{base}}_{\textsc{Mat}}(R)\allowbreak
+\textsc{datamovecost}\!\big(ws^{\textsc{MAT}}(R),x(R)\big)$ &
Memoize (MEM) &
$C_{\textsc{Mem}}(R)=C^{\textsc{base}}_{\textsc{Mem}}(R)+ \textsc{datamovecost}\!\big(ws^{\textsc{MEM}}(R),x(R)\big)$ \\
\bottomrule
\end{tabularx}
\end{table*}

\begin{figure}[h]
  \centering
  \includegraphics[width=0.35\textwidth]{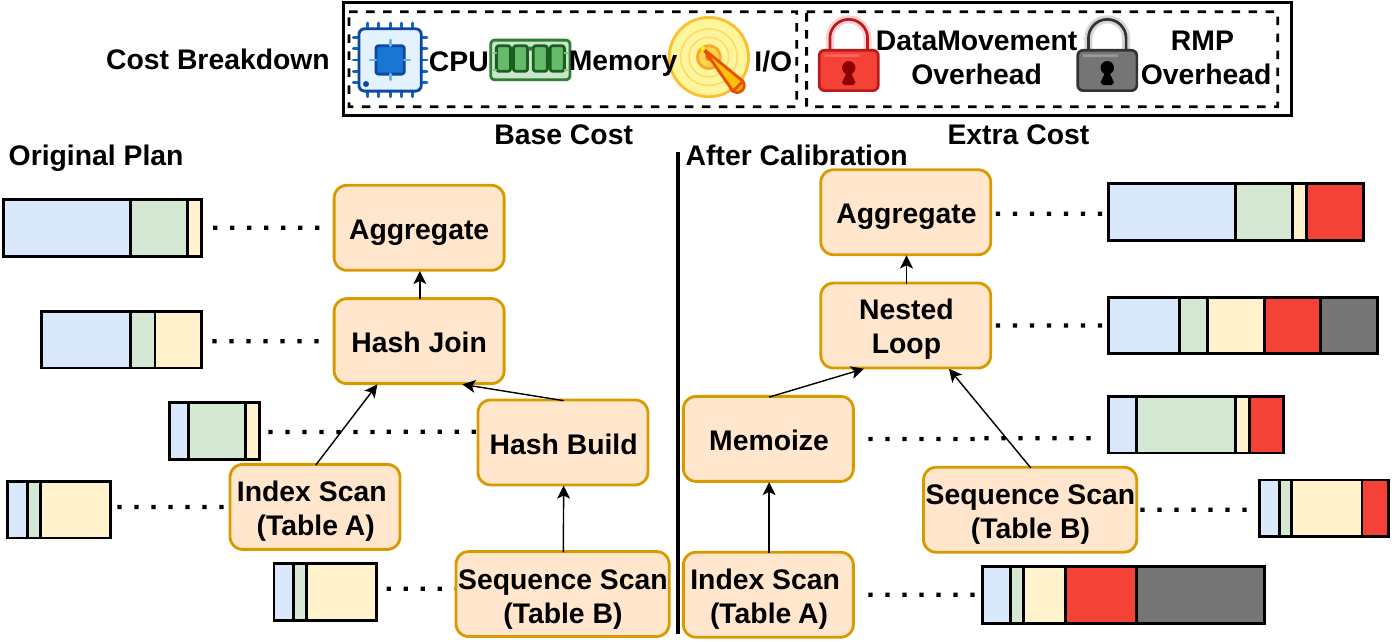}
    \vspace{-5pt}
  \caption{Our light-weight CVM-aware calibration adds two extra overheads for operator cost modeling in CVM.}
  \label{fig:cost_model}
\end{figure}

Cost modeling is the crucial component behind cost-based query optimization: it ranks alternative operators and join orders by predicting execution cost. In CVMs such as SEV-SNP, encryption and integrity mechanisms reshape the relative prices of CPU, memory, and I/O, so conventional cost models can become systematically miscalibrated and prefer CVM-fragile plans. We therefore augment the DBMS's base cost formulas with a small number of CVM-aware terms. As illustrated in Figure~\ref{fig:cost_model}, traditional DBMS cost models decompose operator cost into CPU, memory, and I/O components~\cite{manegold2002generic}, but do not account for the CVM-specific overheads discussed earlier. Our motivation results in Section~\ref{section:motivation} show that these overheads are not uniform across operators. On one hand, different join strategies induce very different bounce-buffer activity patterns, indicating operator-dependent data movement overhead. On the other hand, even under the same SQL and unchanged plan, SEV-SNP still exhibits clear CPI inflation, pointing to additional translation and protected-memory-access cost. These observations motivate our new cost model, summarized in Table~\ref{tab:operator_cost_model}. For each operator, beyond its base cost $C^{\textsc{base}}_{\textsc{Op}}$ capturing conventional CPU and I/O activities, we add CVM-aware overhead terms determined by its input relation $R$ and, when applicable, the other input relation $S$ and index $I$.

\noindent\textbf{Working Set and CVM Cost Primitives.}
A key concept in our model is the \emph{working set} ($ws$), the amount of data that must remain \emph{active} during an operator's execution phase. We derive $ws$ from optimizer-visible statistics: for base-table relations, memory usage is page-oriented, so we approximate $ws$ from \emph{pages} via a byte-level translation; for intermediate results, it is tuple-oriented, so we approximate it as \emph{rows}$\times$\emph{width}. Once $ws$ exceeds cache capacity, more encrypted memory traffic, bounce-buffer interaction, and spill-related movement would be triggered. To capture this excess, we define an overflow factor
\(
x=\max\!\left(0,(ws-(1+\epsilon)C)/C\right),
\)
where $C$ is the effective cache budget and $\epsilon$ is a grace factor introducing a tolerance band near the cache boundary to reduce instability from small estimation errors.
We then model expensive data movement with the \textit{data-movement cost term}
\(
\textsc{datamovecost}(ws,x)=ws\cdot\eta_{\textsc{dm}}\cdot\sigma(x),
\)
where $\eta_{\textsc{dm}}$ is a calibrated per-unit penalty for $ws$ and $\sigma(x)$ activates the term in the overflow regime. The overflow-related parameters and $\eta_{\textsc{dm}}$ should be set so that the term reacts primarily when the active footprint can no longer remain cache-resident. We also model translation and integrity-validation overhead with the \textit{RMP-related cost term}
\(
\textsc{rmptcost}(a,x)=a\cdot\eta_{\textsc{rmp}}\cdot\big(1+\alpha_{\textsc{rmp}}\cdot\sigma(x)\big),
\)
where $a$ captures page-access intensity and locality, $\eta_{\textsc{rmp}}$ is the calibrated per-access penalty for $a$, and $\alpha_{\textsc{rmp}}$ raises the cost under overflow. This follows the second motivation result: even under the same plan, CVM increases cycles and CPI, suggesting the dominant effect is not more work but higher latency along address-translation and protected memory-access paths. We tune these two terms' parameters on a representative subset of queries from each workload; details are in Section~\ref{section:evaluation}. Our goal is not exhaustive optimization, but to show that a reasonable cost-model calibration can reduce the performance gap between KVM and CVM.

\noindent\textbf{Scan Operators.}
\textit{Sequential scans (SS)} exhibit strong spatial locality and predictable streaming access. Their principal CVM vulnerability is therefore not pointer chasing, but the growth of the active scan footprint. When $ws$ exceeds cache capacity, memory traffic increasingly flows through encrypted paths and amplifies bounce-buffer-related overheads. Accordingly, we preserve the base sequential-scan cost and add a $\textsc{datamovecost}$ term. \textit{Bitmap scans and related heap-fetch operators} follow the same logic: their main CVM sensitivity is overflow-driven data movement, so they are also charged via $\textsc{datamovecost}$.

\textit{Index-only and index-driven scans} differ because index-based traversal involves pointer chasing and many small, often non-sequential page accesses. These accesses increase translation intensity and are more aligned with the CPI-inflation evidence in Table~\ref{tab:rmp_evidence_17a}. If a B-tree is used as the index, then the estimated  height is
\(
H(I)=\max\{1,\log_{F}(\textsc{pages}(I))\},
\)
where $F$ denotes the effective fanout implied by the index structure and page layout. We then use $rows_{\textsc{out}}(R)\cdot H(I)$ as the proxy in $\textsc{rmptcost}$, since each returned tuple may trigger a descent across multiple index tree levels. If the accessed data by the index also becomes large, the operator may additionally incur $\textsc{datamovecost}$.

\noindent\textbf{Join Operators.}
Join algorithms expose CVM overheads differently. \textit{Nested-loop joins (NL)} repeatedly probe the inner side for each outer tuple. When the inner side is index-driven, NL can trigger many small and repeated index descents, which raise address translation pressure; hence we add a conditional $\textsc{rmptcost}$ term. Repeated probing and possible buffering enlarge the effective active footprint, so NL also accrues $\textsc{datamovecost}$ when overflow occurs.

\textit{Merge joins (MJ)} are more sequential. Their base behavior is dominated by streaming reads and, if needed, sorting, so their primary CVM exposure is overflow-driven data movement. Still, if one input is produced by an index-heavy plan with weaker locality, translation overhead may also matter, which is why Table~\ref{tab:operator_cost_model} includes a conditional $\textsc{rmptcost}$ term.

\textit{Hash joins (HJ)}'s calibration reflects the clearest working-set phase change. When the build-side hash table fits in the last-level cache (LLC, i.e., the processor's shared L3 cache), performance remains relatively stable; once it exceeds the LLC threshold $W_{L3}$ determined by the system, the operator becomes more sensitive to encrypted memory traffic and random accesses. This behavior is consistent with the motivation evidence that different join strategies induce different data-movement patterns in SEV-SNP. We therefore define
\(
x_{\textsc{HJ}}=\max(0,HT(S)/W_{L3}-1),
\)
where $HT(S)$ is the estimated hash-table size. HJ then incurs both $\textsc{datamovecost}$ for overflow-driven movement and $\textsc{rmptcost}$ for address translation pressure.

\noindent\textbf{Other Operators.}
\textit{Materialize} and \textit{Memoize} operators primarily create or retain intermediate state. Their dominant CVM sensitivity is again governed by the size of the active footprint rather than by complex pointer chasing. Once the buffered state exceeds the cache capacity, encrypted data movement and bandwidth pressure rise. We therefore augment their base costs with $\textsc{datamovecost}(ws,x)$; memoization may also indirectly affect downstream locality by changing later operators' working sets and access patterns.

\section{Preliminary Evaluation}
\label{section:evaluation}

In this section, we evaluate how much time savings the light-weight CVM-aware calibration can achieve and what fraction of queries benefit from it. Finally, we examine the latency distribution before and after applying the cost model calibration.

\noindent\textbf{Experimental Setup.}
We evaluate the calibration on TPC-DS~\cite{tpcds} (scale factor $10$, two instances for each of the 99 query templates), Stack~\cite{marcus2021bao} (10 instances for each of the 16 templates), and IMDB-based~\cite{leis2015good} JOB~\cite{leis2015good} (all 113 queries) and CEB~\cite{flowloss} (10 instances for each of the 16 templates). We set the timeout to $60\text{s}$. We execute the queries in a sequential and reproducible way. For each query, we collect its better performance from enable/disable the calibration. We implemented the calibration in PostgreSQL 16.9. Its default optimizer is our baseline. We use an adapted PostgreSQL~\cite{Bergmann2025Elephant} to support C extension design easily. A default parameter set ($\eta_{\textsc{dm}}$, $\eta_{\textsc{RMP}}$, etc.) is used in all evaluations. Starting from random initial values, we tune one parameter at a time while fixing the others, and retain settings that perform robustly across all workloads using one-third randomly sampled queries per workload. We evaluate the same virtual machine configuration under two security levels: no protection (\textit{KVM}) and \textit{SEV-SNP}. The host machine has 128~GiB RAM and an AMD EPYC~7413 CPU. We configure SEV-SNP/KVM with 64~GiB memory and 32~vCPUs.

\begin{table}[t]
\centering
\caption{The calibration's time savings in SEV-SNP.}
\label{tab:best_system_cache_used}
\small
\begin{tabular}{lccc}
\hline
\textbf{Workload} & \textbf{Baseline (s)} & \textbf{with Calibration (s)} & \textbf{Time Savings} \\
\hline
JOB    & 318.63  & 296.63  & 6.90\%  \\
CEB    & 780.51  & 646.71  & 17.14\% \\
Stack  & 354.72  & 183.20  & 48.35\% \\
TPC-DS & 4049.78 & 3906.28 & 3.54\% \\
\hline
\end{tabular}
\end{table}

\noindent\textbf{Main Evaluation.}
As shown in Table~\ref{tab:best_system_cache_used}, our calibration reduces total workload running time by a range of 3.54\% to 48.35\%. This result highlights a key strength of the CVM-calibrated cost model: calibrating the optimizer's cost model can steer DBMS toward substantially better plans under CVM overhead. For vanilla PostgreSQL in KVM, whose execution times are 259.30\,s (JOB), 677.16\,s (CEB), 329.53\,s (Stack), and 3829.20\,s (TPC-DS), using cost-model calibration in SEV-SNP surpasses the KVM baseline on CEB and Stack, indicating that the calibration effectively bridges a significant portion of the CVM-induced performance gap.

The calibration improves a substantial fraction of queries across all workloads: 43.4\% in JOB, 38.8\% in CEB, 62.5\% in Stack, and 44.4\% in TPC-DS, showing that the benefit of it is broad. The improvement ratio nonetheless varies across workloads, suggesting that workload characteristics affect both how often and how much a fixed calibration helps. One likely reason is that different workloads expose different optimizer weaknesses. For workloads whose performance is highly sensitive to operator choice and relative operator costing, a better CVM-aware cost model can more directly correct plan selection and yield larger savings. In contrast, when the dominant source of error is cardinality estimation, improving operator costs alone has a limited impact. The gap may also reflect differences in plan structure, memory pressure, etc. These observations point to a direction for future work: combining CVM-aware cost calibration with workload-adaptive parameter tuning, and integrating it more tightly with other DBMS components and data properties, has the potential to deliver stronger end-to-end gains.

\begin{figure}
  \centering
  \includegraphics[width=0.36\textwidth]{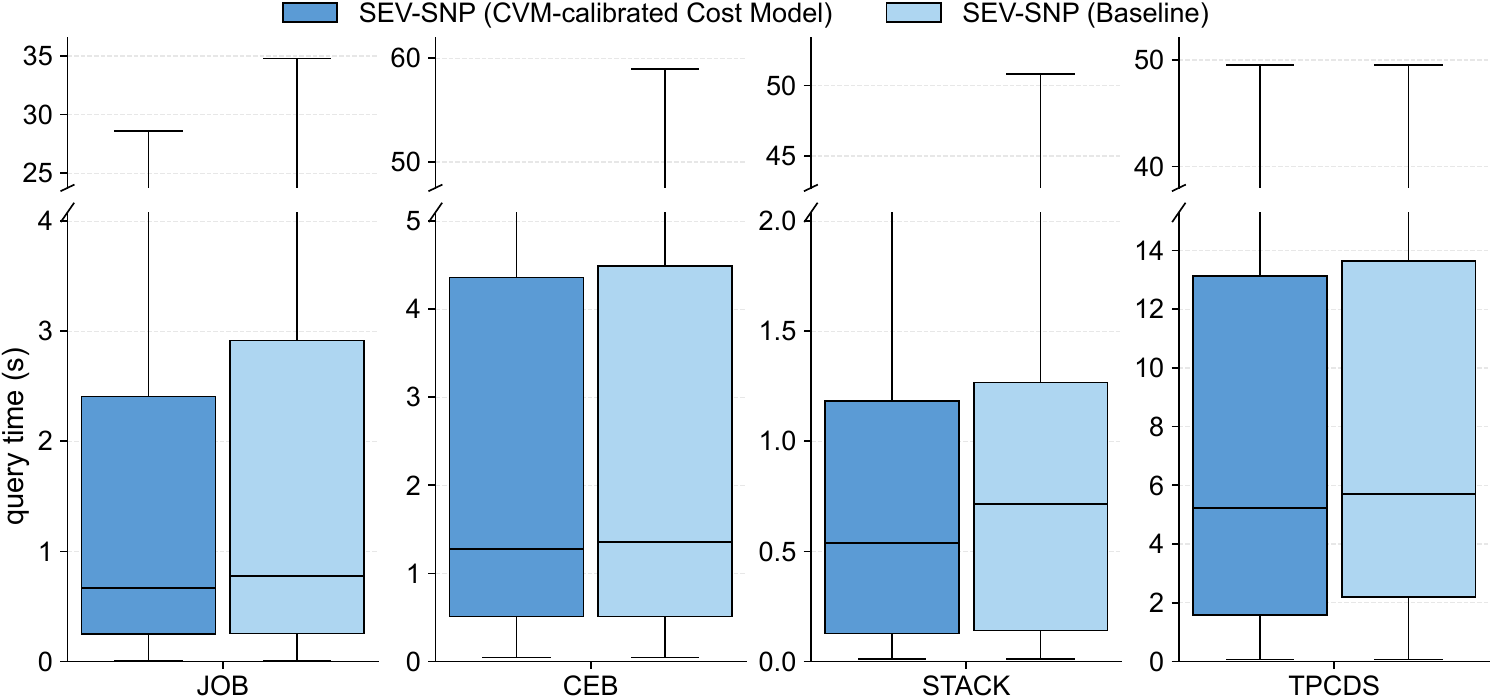}
  \caption{The different query latency distributions.}
  \label{fig:box_plot}
\end{figure}

\noindent\textbf{Latency Distribution.}
We next examine the query latency distribution for each workload in SEV-SNP, comparing vanilla PostgreSQL with the calibration in Figure~\ref{fig:box_plot}. It shifts the distribution downward by lowering the median and compressing the spread. The effect is most striking on Stack, where the median drops from 0.71\,s to 0.53\,s, the standard deviation shrinks from 7.00\,s to 2.24\,s, and the maximum latency falls sharply from 50.80\,s to 13.84\,s. CEB shows a similar long-tail reduction, with the maximum 28.12\,s decrease. JOB also improves more broadly, with both median latency reduced (0.77\,s to 0.66\,s) and the tail shortened (34.77\,s to 28.60\,s). The calibration corrects the optimizer's cost reasoning under CVM overhead and helps PostgreSQL avoid fragile plans in SEV-SNP.




\section{Related Work}
\label{section:related_work}
Prior work on DBMSs under TEEs mainly spans four directions. First, existing TEE-based DBMSs~\cite{eskandarian2017oblidb,vinayagamurthy2019stealthdb,fuhry2021encdbdb,li2023encrypted,han2021prodb} focus on security architecture and trusted execution. Second, operator-level studies design oblivious joins, sorting, and related primitives with formal leakage protection~\cite{krastnikov2020efficient,maliszewski2023cracking,gu2025flexway,mavrogiannakis2025obliviator}. Third, benchmarking work shows that trusted execution can substantially slow analytical workloads~\cite{maliszewski2021price,battiston2024duckdb,lutsch2024benchmarking,lutsch2025analysis}; more broadly, a growing line of work benchmarks CVMs against KVM baselines and studies how to optimize CVM execution beyond DBMSs~\cite{yan2023performance,misono2024confidential,holmes2024severifast}. Finally, our work also relates to database research on adapting cost models to hardware characteristics~\cite{bausch2012making,manegold2002generic}. Unlike prior TEE database work, which mainly emphasizes secure execution or overhead measurement, we target the underexplored optimizer side: improving database performance in CVMs through lightweight cost-model calibration.

\section{Conclusion}

We present a light-weight CVM-aware calibration for the cost model and integrate it into PostgreSQL. 
Our design augments a default cost modeling in the query optimizer with simple penalties that better reflect CVM-specific overheads: data movement and RMP checks. This calibration can improve query planning decisions and reduce execution time in SEV-SNP, which mitigates the performance gap between KVMs and CVMs.

\bibliographystyle{plain}
\bibliography{references}
\end{document}